\documentclass[aps,twocolumn,pra]{revtex4}
\usepackage{amsmath}
\usepackage{graphicx}
\usepackage{epsfig}
\usepackage{bm}

\newcommand{\be}{\begin{equation}}
\newcommand{\ee}{\end{equation}}

\newcommand{\beq}{\begin{eqnarray}}
\newcommand{\eeq}{\end{eqnarray}}

\def\H1{\widehat{H}_1}

\begin{document}

\title{Supplementary Information to the paper ``Breakdown of the adiabatic limit in low dimensional gapless systems''}

\author{Anatoli Polkovnikov$^1$ and Vladimir Gritsev$^2$}
\affiliation {$^1$Department of Physics, Boston University, Boston, MA 02215\\
$^2$Department of Physics, Harvard University, Cambridge, MA 02138 }

\begin{abstract}
Here we give detailed derivations and provide additional examples to
the main paper~\cite{main}. In particular, we discuss the scaling behavior of observables like correlation functions and density of excitations.
We also analyze effects of nonintegrability of the Bose-Hubbard
model on the long-time dynamics of the correlation functions. In
addition we explicitly consider several interacting models, where we
are able to analyze slow dynamics and classify it according to the
regimes suggested in the main paper.
\end{abstract}

\maketitle

\section{Introduction and Discussion}
The aim of this Supplementary Material is twofold. First, in
Sections II-V we provide more details to the main text~\cite{main}. Thus in
Sec.~\ref{APP:Pert} we describe the approach to the slow dynamics in
our model based on the Fermi Golden Rule in the ramping rate. In
Sec.~\ref{App:A} we give details of the time evolution of the wave
function of the harmonic model if the system is initially prepared
in the ground state. Then in Sec.~\ref{App:B} we generalize this
derivation to the evolution of the density matrix assuming the
initial thermal distribution. Dynamics of the Bose-Hubbard model is
addressed in Sec.~\ref{App:C}. There we give the details of our
numerical approach based on the semiclassical approximation and the
leading quantum corrections. We also discuss the consequences of
non-integrability of the Bose-Hubbard model on the time evolution of
the correlation functions. In particular, we show that at very long
times the non-equilibrium state created during the ramp relaxes to
the thermal equilibrium.

In the main paper~\cite{main} we exclusively concentrated on finding energy
added to the system during the ramp. In this Supplementary Material
we will also consider various other quantities like density of
excitations, correlation functions etc. Sometimes these quantities
are easier to measure experimentally and if the effects of
non-integrability are weak or absent then they are good observables
to work with.

The second aim of this Supplementary Material is to consider
application of our findings to several physical problems in more
details. In Sec.~\ref{appl} we discuss potential relevance of our
findings to the quantum information and to some problems in
inflationary cosmology. Then in Sec.~\ref{TG} discuss the dynamics
of one dimensional bosons in Tonks-Girardeau regime~\cite{KWW}. In
particular, we show that if the trapping potential for atoms is
slowly reduced to zero then the heating induced in the system is
described by the non-analytic {\bf B}) regime according to our
classification. We show that these results can also be applied to
the Calogero-Sutherland model~\cite{CS}, describing one dimensional
fermions with long range interactions, in the harmonic trap. And
finally in Sec.~\ref{Dicke} we briefly describe time evolution  of
the quantum Dicke model~\cite{Dicke-model}, which serves as a
prototype for the laser, as well as mimics a coherent atomic cloud
in the cavity QED. There in a particular regime we find that the
number of generated photons is described by the non-adiabatic regime
{\bf C}). Of course this list of possible applications of our
findings is not complete, yet it is quite illuminating. We note that
some of these models are integrable, some are not. Yet we find
general good agreement of the slow dynamics in all of these models
with our classification scheme.

\section{Fermi Golden Rule analysis of the slow dynamics.}
\label{APP:Pert}

One way to find the density of excitations $n_{\rm ex}$ and the
energy produced during a slow increase is to use the Fermi Golden in
the ramp speed~\cite{adiabatic}. We remind that we consider the
Hamiltonian~(5) from the main text:
\be
\mathcal H=\sum_q {\rho_s q^2\over 2}|\phi_q|^2+{1\over
2}\kappa_q |\Pi_q|^2,
\label{Hamiltonian}
\ee
where we choose $\kappa_q=\kappa+\lambda q^2$ and
$\kappa=\kappa_0+\delta t$ linearly change in time. Since $n_{\rm
ex}$ should be small at small $\delta$ one can expect that the
perturbation theory in $\delta$ gives a good estimate of $n_{\rm
ex}(\delta)$. Then the density of excitations $n_{\rm ex}$ can be
expressed as follows~\cite{adiabatic}:
\begin{widetext}
\be
n_{ex}\approx {1\over L^d}\sum_m^\prime
\left|\int\limits_{\kappa_0}^\infty d\kappa \left< m\left|{d\over
d\kappa} \right|0\right>_\kappa\, \exp\left[ {i\over\delta}
\int\limits_{\kappa_0}^\kappa
(\omega_m(\kappa^\prime)-\omega_0(\kappa^\prime))d\kappa^\prime\right]
\right|^2,
\label{eq:2}
\ee
\end{widetext}
where $|m\rangle_\kappa$ denotes a general excited many-body state
with the energy $\hbar\omega_m(\kappa)$, $\langle m|d/d\kappa
|0\rangle_\kappa$ is the matrix element of the derivative with
respect to $\kappa$ between the states $|m\rangle$ and $|0\rangle$
at a given value of $\kappa$. We would like to emphasize that
Eq.~(\ref{eq:2}) is valid only if the ground state evolution does
not acquire an additional Berry phase. In the situation where the
Berry phase is nonzero it should be subtracted from the argument of
the exponent in this equation. It is straightforward to check that
with the Hamiltonian~(\ref{Hamiltonian}) the only non-vanishing
matrix element of $d/d\kappa$ corresponds to the excitation of two
quasiparticles with opposite momenta: $\langle {\bf q},{\bf -q}|
\partial_\kappa|0\rangle=1/(4\sqrt{2}\kappa_q)$.
Using that $\omega_{m}-\omega_0=2q\sqrt{\rho_s \kappa_q}$,  where
the factor of two comes from the fact that we have two excited quasi-particles, we find:
\be
n_{ex}={1\over 32}\int {d^d q\over
(2\pi)^d}\left|\int_{\kappa_{0,q}}^\infty {d\xi\over\xi}\exp\left(
{4i\over 3\delta}\sqrt{\rho_s}q\,\xi^{3/2}\right)\right|^2,
\label{eq:3}
\ee
where $\kappa_{0,\,q}=\kappa_0+\lambda q^2$. This expression gives
different asymptotics in the two opposite limits.

(i) If $\delta\gg \kappa_0^2\sqrt{\rho_s/\lambda}$, which is the
case if one starts from the weakly interacting regime $\kappa_0\to
0$, then
\be
n_{ex}\approx A_d {\delta^{d/4}\over
\rho_s^{d/8}\lambda_{\phantom{s}}^{3d/8}},
\label{eq:7}
\ee
where $A_d$ is a numerical constant. It is easy to check that if
$d>d^{\star}=8$ the exponent of $\delta$ saturates at $2$ and does
not depend on the dimensionality. Expression (\ref{eq:7}) suggests
that in this particular situation the nonanalytic regime {\bf B} is
realized in all physical dimensions. In one dimension it is
particularly hard to reach the adiabatic regime since $n_{\rm ex}$
scales only as $\delta^{1/4}$.

We note that the scaling in Eq.~(\ref{eq:7}) is consistent with the
one obtained in Ref.~[\onlinecite{adiabatic}] for the crossing of
the second order phase transition: $n_{\rm ex}\propto
\delta^{d\nu/(z\nu+1)}$, where $\nu$ is the critical exponent
characterizing divergence of the correlation length. In our case
there is a diverging healing length $\xi\sim \sqrt{\lambda/\kappa}$
instead of the correlation length (see Ref.~[\onlinecite{ps}] for
details) so that $\nu=1/2$ and given that $z=2$ in the
noninteracting regime one immediately recovers that
$\nu/(z\nu+1)=1/4$.

(ii) In the opposite limit, where the initial value of $\kappa$ is
large $\delta\ll\kappa_0 \sqrt{\rho_s/\lambda}$ the situation
becomes more diverse. Thus for dimensions $d<2$ Eq.~(\ref{eq:7})
yields
\be
n_{ex}\approx A_d^\prime {\delta^d\over
\rho_s^{d/2}\kappa_0^{3d/2}}.
\label{n_ex}
\ee
On the other hand for $d>2$ the exponent saturates and we have
\be
n_{ex}\approx A_d^\prime {\lambda^{1-d/2}\kappa_0^d\over
\rho_s}\,\delta^2.
\label{eq:11}
\ee
In two dimensions there is an additional logarithmic correction to
the scaling (\ref{n_ex}). We see that in this situation the critical
dimension above which the analytic regime holds is $d^\star=2$.

The present analysis can be generalized to other situations. For
example, in the case of ferromagnets $\kappa_0\equiv 0$ and then one
can tune $\lambda$. Then one finds that $n_{\rm ex}\propto
\delta^{d/2}$ and the critical dimension is $d^\star=4$. We comment
that one can also consider other scenarios of varying $\kappa$ with
time. For example, if $\kappa\propto (\delta t)^r$ then it is easy
to see that $n_{\rm ex}\propto \delta^{dr/2(r+1)}$. As $r$ increases
the scaling of the density of excitations interpolates from
$\delta^{d/4}$ to $\delta^{d/2}$ and changes $d^\star$ from eight to
four consistent with a recent prediction of Ref.~\cite{krishnendu}.

This perturbative analysis shows the existence of {\bf A} ({\bf B})
regimes for dimensions above (below) some critical value $d^\star$.
However, it misses the existence of the non-adiabatic {\bf C} regime.
To justify the validity of the application of the Fermi golden rule
one has to require that the probability of excitation of each
momentum mode is small. This requirement breaks down at low energies
as can be readily seen from Eq.~(\ref{eq:3}). In the case when the
excitations have Fermionic character, which is e.g. the case for
crossing the critical point in the transverse field Ising model or
the XXZ chain~\cite{subir}, the mistake of the perturbative
treatment is a simple factor of the order of one (see
Refs.~[\onlinecite{adiabatic, zurek1, jacek, fubini}]). The
Goldstone modes described by the Hamiltonian~(\ref{Hamiltonian})
are harmonic oscillators and thus behave as bosons. Bosons unlike
fermions have a bunching tendency, i. e. transition probabilities
can be significantly enhanced compared to the golden rule
prediction.

For the energy density in the system one can derive a similar expression to Eq.~(\ref{eq:3}):
\be
\mathcal E={1\over 32}\int {d^d q\over
(2\pi)^d}\sqrt{\kappa_f\rho_s}q\left|\int_{\kappa_{0,q}}^\infty {d\xi\over\xi}\exp\left(
{4i\over 3\delta}\sqrt{\rho_s}q\,\xi^{3/2}\right)\right|^2.
\label{eq:en_pert}
\ee
From here, for example, for the initially noninteracting case $\kappa_0=0$ one recovers Eq.~(6) of the main text:
\be
\mathcal E\propto {\delta^{(d+1)/4}\over
(\rho_s\lambda^3)^{(d+1)/8}}.
\label{en_0}
\ee
Similarly one can reproduce the correct scaling for finite $\kappa_0$ mentioned in the main text.

\section{Evolution of the wave function at zero initial temperature.}
\label{App:A}
The harmonic theory described by the Hamiltonian~(\ref{Hamiltonian})
can be in principle analyzed for arbitrary functional dependence of
coupling $\kappa_q(t)$. This is a consequence of the fact that the
quantum harmonic oscillator problem can be solved for arbitrary
functional dependence of its frequency (and mass) on time. The
resulting Riccati-type equation can be analytically solved in some
situations. In particular, this is the case for the linear time
dependence, which we analyze here in more detail.

As we described in the main text the initial ground state wave function is given by
\be
\Psi(\{\phi_q\})=\prod_q {1\over
(2\pi\sigma_{0,\,q})^{1/4}}\exp\left[-{|\phi_q|^2\over
4\,\sigma_{0,\,q}}\right],
\label{psi}
\ee
where $ \sigma_{0,\,q}=1/ (2q)\sqrt{\kappa_{0,q}/\rho_s }$. If
$\kappa$ changes with time, $\sigma_q$ acquires time dependence:
\be
i {d \sigma_q\over dt}=2\rho_s q^2\sigma_q^2-{1\over 2}\kappa_q(t).
\label{eq:16}
\ee
This equation can be simplified by first changing independent
variable $t$ to $\kappa_q(t)$ and then by a simple rescaling:
\be
\kappa=\tilde\kappa {\delta^{2/3}\over \sqrt[3]{\rho_s q^2}},\;
\sigma_q=\tilde\sigma_q {\delta^{1/3}\over 2\sqrt[3]{\rho_s^2
q^4}},\; q=\tilde q {\delta^{1/4}\over \rho_s^{1/8}\lambda^{3/8}}.
\label{eq:18a}
\ee
Under these transformations we also have $\tilde
\kappa_q=\tilde\kappa+\tilde q^{8/3}$. Then one can check that
Eq.~(\ref{eq:16}) is equivalent to
\be
i{d\tilde \sigma_q\over d\tilde \kappa_q}=\tilde \sigma_q^2-\tilde
\kappa_q.
\label{eq:18}
\ee
This Riccati equation can be explicitly solved in terms of Airy
functions ${\rm Ai}$ and ${\rm Bi}$:
\be
\tilde \sigma_q=-i {{\rm Bi}^\prime (-\tilde \kappa_q)+\alpha_q {\rm
Ai}^\prime (-\tilde \kappa_q)\over {\rm Bi}(-\tilde
\kappa_q)+\alpha_q {\rm Ai}(-\tilde \kappa_q)},
\ee
where $\alpha_q$ is an integration constant, which is determined
from the initial conditions. In the limit $\tilde \kappa_q\to
\infty$ ignoring unimportant fast oscillating terms we find
\be
\Re \left[1\over \tilde\sigma_q\right]\to {2\Im\alpha_q\over
\sqrt{\tilde \kappa_q}[1+|\alpha_q|^2]}.
\ee
Note that the real part of $1/\sigma_q$ determines $|\psi|^2$ and
thus the probability distribution of the corresponding Fourier
component of the phase $\phi_q$ (see Eq.~(\ref{psi})). The
fact that $1/\sigma_q\to 0$ as $\kappa_q\to\infty$ should not be
surprising. Indeed the width of the ground state wave function in
scaled variables is
\be
\tilde{\sigma}^{\rm eq}_q=\sqrt{\tilde \kappa_q}\approx {1\over
2q}\sqrt{\kappa\over\rho_s}\,\left({ 2\sqrt[3]{\rho_s^2 q^4}\over \delta^{1/3}}\right).
\label{sigma_eq}
\ee
The probability of excitations in the system is determined by the
ratio of $\sigma_q$ and $\sigma^{\rm eq}$, which takes a well
defined limit at $\kappa\to\infty$. Introducing $\sigma^{\rm
eff}_q=1/\Re(\sigma_q^{-1})$ we find
\be
{\sigma^{\rm eff}_q\over\sigma^{\rm eq}_q}={1+|\alpha_q|^2\over
2\Im\alpha_q}.
\label{sig_eff}
\ee
The initial condition determining $\alpha$ is:
\be
\sqrt{\tilde \kappa_{0,\,q}}=i { {\rm Bi}^\prime(-\tilde
\kappa_{0,\,q})+\alpha_q {\rm Ai}^\prime(-\tilde
\kappa_{0,\,q})\over {\rm Bi}(-\tilde \kappa_{0\,q})+\alpha_q {\rm
Ai}(-\tilde \kappa_{0,\,q})}.
\ee
This equation can be inverted to give
\be
\alpha_q=-{\sqrt{\tilde \kappa_{0,\,q}}\, {\rm Bi}(-\tilde
\kappa_{0,\,q})-i {\rm Bi}^\prime(-\tilde \kappa_{0,\,q})\over
\sqrt{\tilde \kappa_{0,\,q}}\, {\rm Ai}(-\tilde \kappa_{0,\,q})-i {\rm
Ai}^\prime(-\tilde \kappa_{0,\,q})}.
\label{alpha_q}
\ee

In the limit $\tilde\kappa_{0,\,q}\ll 1$ this equation yields:
\be
\alpha_q\approx \sqrt{3}+i{3^{2/3}\Gamma^2(1/3)\over
\pi}\sqrt{\tilde \kappa_{0,\,q}}.
\ee
Consequently
\be
{\sigma^{\rm eff}_q\over\sigma^{\rm eq}_q}\approx {2\pi\over
3^{2/3}\Gamma^2(1/3)}{1\over \sqrt{\tilde \kappa_{0,\,q}}}.
\label{eq:26}
\ee
In the opposite limit $\tilde\kappa_0^q\gg 1$ one finds
$\alpha_q\approx i$ and
\be
{\sigma^{\rm eff}_q\over\sigma^{\rm eq}_q}\approx 1+{1\over 32
\tilde \kappa_{0,\,q}^{\,3}}.
\label{eq:28}
\ee
We note that in this limit Eq.~(\ref{eq:28}) gives the result
identical to what one would get using Fermi Golden rule approach
described in the previous section:
\be
n_q={1\over 144}\left|\Gamma\left(0,-i{4\over 3}\tilde
z_0^{3/2}\right)\right|^2\approx {1\over 64}{1\over \tilde \kappa_{0,\,q}^3}.
\label{n_q_pert}
\ee
We remind that $n_q$ and $\sigma_q$ are related according to Eq.~(13) of the main text:
\be
n_q={1\over 2}\left[{\sigma^{\rm eff}_q\over\sigma^{\rm eq}_q}-1\right]
\ee
so that Eqs.~(\ref{eq:28}) and (\ref{n_q_pert}) indeed agree for high energy modes.

The number of excitations studied above is not necessarily an
observable quantity. Instead one can look, for example, into the
behavior of the correlation functions, which are closely related to the population of different modes:
\be
\left<\mathrm e^{i(\phi(x)-\phi(0))}\right>=\exp\left[-\sum_{q\neq 0}
\sigma_{\rm eff}^q \sin^2 qx/2 \right].
\label{corr_func}
\ee

If the initial state is noninteracting: $\kappa_0=0$ then according
to Eqs.~(\ref{eq:26}), (\ref{sigma_eq}), and (\ref{eq:18a}) we have
$\sigma_{\rm eff}^q\propto q^{-7/3}$ at small $q$. Therefore in one
and two dimensions
\be
\left<\mathrm e^{i(\phi(x)-\phi(0))}\right>\sim \exp [-C
\delta^{1/3} x^{7/3-d}].
\label{eq:30}
\ee
In one dimension this integral decays faster than exponential
indicating that the system is overheated, i.e. the behavior of the
correlation functions is worse than at finite temperature. In two
dimensions the correlation functions decay as $\exp[-C\delta^{1/3}
x^{1/3}]$, which is again a very unusual behavior. Note that in one
and two dimensions the asymptotic behavior of correlation functions
(\ref{eq:30}) is valid only at long distances $x\gtrsim \xi_{d}$
with $\xi_{1D}\sim 1/\delta^{1/4}$ and $\xi_{2D}\sim 1/\delta$. In
dimensions $d>7/3$ the excitations in the system do not destroy the
long-range order in the system but reduce the superfluid density
\be
\lim_{{\bf r}\to\infty}\left<\mathrm e^{i(\phi({\bf
r})-\phi(0))}\right>\sim \exp[-A\,\delta^{(d-1)/4}].
\ee

If one starts in the interacting regime:
$\kappa_0\gg\sqrt{\delta/n_0}$ then one finds that $\sigma_{\rm
eff}(q)\propto 1/q^{4/3}$ at small $q$ and thus the correlation
functions are singular only in one dimension:
\be
\left<\mathrm e^{i(\phi(x)-\phi(0))}\right>\sim \exp [-C
\delta^{1/3} x^{1/3}].
\ee
In this case the correlation length diverges as $\xi_{1D}\sim
1/\delta$. Above one dimension the long range order survives and the
long distance behavior of correlation functions is:
\be
\lim_{{\bf r}\to\infty}\left<\mathrm e^{i(\phi({\bf
r})-\phi(0))}\right>\sim \exp[-\tilde A\,\delta^{d-1}]
\ee
for $d<3$. Above three dimensions the power of $\delta$ in the
expression above saturates at two.

We would like to stress that the steady state non-equilibrium
distribution of quasi-particles and as a consequence noneqiulibrium
correlation functions, which we obtained above are only possible in
strictly noninteracting model. Indeed addition of small nonlinear
terms into the Hamiltonian~(\ref{Hamiltonian}) can lead to
redistribution of excitations among different states and eventual
thermalization. We already highlighted that this is indeed the case
in the main text and will return to this issue again in Sec.~\ref{App:C}. We note that these possible thermalization processes do not affect the total energy, which is conserved in an isolated system.

\section{Evolution of the density matrix at finite initial temperature.}
\label{App:B}

We choose to represent the density matrix corresponding to the
initial thermal state in the Wigner form~\cite{gardiner, walls}. For
the harmonic system described by the Hamiltonian~(5) one can show
that this density matrix factorizes into the product of Gaussian
functions:
\be
W_0=\prod_q {1\over 2\pi r_q}\exp\left[-{|\phi_{0,\, q}|^2\over
2\sigma_{0,\,q}r_q}-{\sigma_{0,\,q}|\Pi_{0,q}|^2\over 2 r_q}\right],
\label{eq:45}
\ee
where
\be
r_q=\coth\left[{q\sqrt{\kappa_{0,\,q} \rho_s}\over 2 T}\right].
\ee
In the noninteracting problem the time
evolution of the fields $\phi_{\bf q}$ and $\Pi_{\bf q}$ is
described by the classical equations of motion~\cite{gardiner, twa}:
\be
{d\over dt}\left[{1\over \kappa_q}{d\phi_q\over dt}\right]+\rho_s
q^2\phi=0,
\label{eq:47}
\ee
subject to the initial conditions
\be
\phi_q(t=0)=\phi_{0,\, q},\; \dot\phi_q(t=0)=\kappa_{0\,q} \Pi_{0,\,
q}.
\ee
Here $\phi_{0, q}$ and $\Pi_{0, q}$ are randomly distributed
according to Eq.~(\ref{eq:45}). The other important feature of Gaussian
ensembles is that in the absence of interactions the Wigner
distribution (\ref{eq:45}) always preserves its Gaussian form.
Therefore finding $\langle \phi_q^2(t) \rangle$ and $\langle
\Pi_q^2(t)\rangle$ is sufficient to fix the whole distribution
function at arbitrary time. Alternatively one can directly solve
the Liouville equation for the density matrix in the Wigner
form~\cite{gardiner} and come to the same conclusion.

A general solution of Eq.~(\ref{eq:47}) is:
\be
\phi_q(\tilde \kappa_q)=C_1 {\rm Ai}^\prime(-\tilde
\kappa_q)+C_2{\rm Bi}^\prime (-\tilde \kappa_q),
\ee
where as in Appendix~\ref{App:A} we changed the variables from $t$
to $\tilde \kappa_q$. The integration constants $C_1$ and $C_2$ can
be found from the initial conditions:
\beq
&&C_1={\pi \kappa_{0,\,q}\over \tilde
\kappa_{0,\,q}^{\,2}}{d\phi_{0,\,q}\over d\kappa_{0,\,q}}{\rm
Bi}^\prime(-\tilde \kappa_{0,\, q})
-\pi\phi_{0,\,q} {\rm Bi}(-\tilde \kappa_{0,\,q}),\phantom{XX}\\
&&C_2=\pi\phi_{0,\,q} {\rm Ai}(-\tilde \kappa_{0,\,q})-{\pi
\kappa_{0,\,q}\over \tilde \kappa_{0,\, q}^{\,2}}{d\phi_{0,\,q}\over
d\kappa_{0,\, q}}{\rm Ai}^\prime(-\tilde \kappa_{0,\,q}).
\eeq
From these expressions it is easy to find the asymptotical behavior
of $\langle \phi_q^2\rangle$ at large $\tilde \kappa$ and thus find
the width of the distribution $\sigma^{\rm eff}_q$:
\beq
&&{\sigma^{\rm eff}_q\over \sigma^{\rm eq}_q}={\pi\over 2} {r_q\over
\sqrt{\tilde \kappa_{0,\, q}}}\bigl[\tilde \kappa_{0,\,q}{\rm
Bi}^2(-\tilde \kappa_{0,\,q})+\tilde
\kappa_{0,\,q}{\rm Ai}^2(-\tilde \kappa_{0,\, q})\nonumber\\
&&~~~~~~~~~~~~~~~~~+({\rm Bi}^\prime(-\tilde \kappa_{0,\,
q}))^2+({\rm Ai}^\prime(-\tilde \kappa_{0,\,q}))^2\bigr].
\eeq
One can verify that apart from the factor of $r_q$, which approaches
unity at $T\to 0$, the expression above coincides with the zero
temperature results (see Eqs.~(\ref{sig_eff}) and (\ref{alpha_q})), i.e.
\be
{\sigma^{\rm eff}_q(T)\over \sigma^{\rm eq}_q}=r_q(T) {\sigma^{\rm eff}_q(T=0)\over \sigma^{\rm eq}_q}.
\label{sig_eff_temp}
\ee
This result immediately implies that the number of the additional
excitations created during the ramp at finite temperature can be
obtained from the zero temperature result by multiplication by $r_q$:
\be
n_q={1\over 2}\left[{\sigma^{\rm eff}_q\over \sigma^{\rm
eq}_q}-1\right]=r_q \left.n_q\right|_{T=0}+{1\over 2}(r_q-1).
\label{n_qt}
\ee

Integrating $n_q$ over momenta we find that the total density of
excitations for $\kappa_0=0$ scales at finite temperature in all
three spatial dimensions as
\be
n_{\rm ex}\propto T L^{10/3-d}\sqrt[3]{\delta}.
\ee
So in terms of the excitation density the system is always in the
regime {\bf C}). The energy density shows less divergent behavior
and the regime {\bf C}) is realized only in one and two spatial
dimensions while in the three dimensional case dynamics belongs to
the nonanalytic {\bf B}) as described in the main text~\cite{main}.

If the initial compressibility $\kappa_0$ is finite than we find
that in one dimension the density of excitations still diverges with
the system size:
\be
n_{\rm ex}\propto T\sqrt[3]{\delta L},
\ee
but it is finite in two and three dimensions $n_{\rm ex}\propto
T\delta^{d-1}$ (as before the exponent of $\delta$ saturates at two
for $d>3$). The total energy converges in all three dimensions and
it behaves as $\mathcal E_{\rm f}\propto T\delta^d$ for $d< 2$ and
$\mathcal E_{\rm f}\propto T\delta^2$ for $d>2$.

As in the previous section one can compute correlation functions.
Note that because for initially noninteracting regime $\sigma_q$
diverges as $1/q^{13/3}$ the sum in Eq.~(\ref{corr_func}) is
infrared divergent in one and two dimensions even at $qx<1$. This
results in a very unusual behavior of the correlation functions.
\be
\left<\mathrm e^{i(\phi(x)-\phi(0))}\right>\sim \exp\left[-C T\sqrt[3]{\delta}\, x^2L^{7/3-d}\right].
\label{corr_func1}
\ee
In three dimensions we have
\be
\left<\mathrm e^{i(\phi(x)-\phi(0))}\right>\sim \exp\left[-C T\sqrt[3]{\delta}\, x^{4/3}\right].
\ee
We comment again that this unconventional behavior of the
correlation functions can exist as long as the relaxation processes
in the system are negligible. We will get back to this issue in the
next section.

\section{Quantum dynamics of a Bose-Hubbard system: expansion in quantum fluctuations.}
\label{App:C}

Here we will describe in some detail how  to simulate slow dynamics
of the system described by the Hubbard model~(11) of the main text
using the semiclassical approach~\cite{twa}. For completeness we
will write the Bose-Huabbard Hamiltonian again:
\be
\mathcal H_{bh}=-J\sum_{\langle ij\rangle} (a_i^\dagger
a_j+a_j^\dagger a_i)+ {U(t)\over 2}\sum_j a_j^\dagger
a_j(a_j^\dagger a_j-1),
\label{h_bh}
\ee
Here $a_j$ and $a_j^\dagger$ are the bosonic annihilation and
creation operators, $J$ represents the tunneling matrix element and
$U$ is the interactions strength. The sum in the first term is taken
over the nearest neighbor pairs.

Specifically we will use expansion of the time evolution of the
system in the small quantum parameter $U/Jn_0$.  Note that when this
parameter is close to one, the ground state of the system undergoes
the superfluid-insulator transition driven by quantum
fluctuations~\cite{subir}. Conversely when $U/Jn_0\ll 1$ quantum
fluctuations are negligible and the system is in the superfluid
regime. From this one can conclude that this ratio plays the role of
the Planck's constant in this problem (see Ref.~\cite{psg} for more
details). Here we are interested in evolution precisely in the
regime where the system is far from the insulating phase and the
harmonic approximation is accurate so the expansion in this ratio is
justified.

We note for those more familiar with the Keldysh
technique~\cite{kamenev} that our approach treats all classical
vertexes exactly and expands the evolution in number of quantum
vertexes. In the leading order in this parameter one obtains the so
called truncated Wigner approximation (TWA)~\cite{walls, steel},
where the classical fields $\psi_j^\star$ and $\psi_j$ corresponding
to the operators $a_j^\dagger$ and $a_j$ satisfy the time dependent
Gross-Pitaevskii equations of motion. In the next order the
classical fields are subject to a single quantum jump during the
evolution. We find that while TWA approximation is adequate at
finite temperatures, in the zero temperature limit one has to go
beyond and add the next correction. This finding agrees with a
general statement that the semiclassical approximation can break
down at long times~\cite{twa, psg}.

In the classical limit the bosonic fields $\psi_j^\star$ and
$\psi_j$ satisfy the discrete Gross-Pitaevskii equations:
\be
i{\partial \psi_j\over \partial t}=-J\sum_{i\in O_j} \psi_i+U(t)
|\psi_j^2|\psi_j.
\ee
Here the sum in the first term is taken over the nearest neighbors
of the site $j$. In the leading order in quantum fluctuations, which
corresponds to the semiclassical or truncated Wigner approximation
(TWA), the fields $\psi_j$ and $\psi_j^\star$ are subject to random
initial conditions, which are distributed according to the Wigner
transform of the initial density matrix $W(\psi_j^\star,\psi_j)$.
The expectation value of an arbitrary observable
$\Omega(a_j^\dagger, a_j)$ is given by the average of the
corresponding Weyl symbol (fully symmetrized form of the operator)
$\Omega_{\rm cl}(\psi_j^\star,\psi_j)$ on the solutions of the
Gross-Pitaevskii equations:

\be
\langle \Omega(t)\rangle_0=\int D\psi^\star_jD\psi_j
W(\psi_j^\star,\psi_j)\Omega_{\rm cl}(\psi_j^\star(t)\psi_j(t)).
\label{eq:twa}
\ee

Since the initial system is noninteracting, it is straightforward to
find the Wigner transform of the density matrix at finite
temperature $T$. It is more convenient to write it in the Fourier
space
\be
W(\hat\psi_k^\star,\hat\psi_k)=Z\prod_q
\exp\left[-2|\hat\psi_q|^2\tanh\left(\epsilon_0(q)-\mu\over
2T\right)\right],
\ee
where $\hat\psi_k$ is the discrete Fourier transform of $\psi_j$,
$Z$ is the normalization constant, $\epsilon_0(q)=-J\sum_{j}\mathrm
e^{iqj}$ is the excitation energy of the Bose-Hubbard
Hamiltonian~(\ref{h_bh}) in the absence of interactions and the
summation is taken over nearest neighbors of site at the origin,
$\mu$ is the chemical potential which enforces mean number of
particles per site $n_0$. We note that in large systems we consider
here, there is no difference in time evolution between grand
canonical and canonical ensembles~\cite{ap_cat}.

We find that the semiclassical approximation (\ref{eq:twa}) gives
very accurate results in most of our simulations described in this
paper. However, at zero temperature case it breaks down for very
slow ramps and we had to include the next quantum correction to the
TWA. The latter manifests itself in the form of a single
infinitesimal quantum jump during the evolution:
\be
\psi_i(t^\prime)\to \psi_i(t^\prime)+\epsilon_1+i\epsilon_2.
\ee
The quantum correction is the evaluated as a nonlinear response of
$\Omega_{\rm cl}$ to such a jump~\cite{twa}:
\begin{widetext}
\be
\langle \Omega(t)\rangle_1=-\int D\psi^\star_jD\psi_j
W(\psi_j^\star,\psi_j)\sum_i \int_0^t dt^\prime {U(t^\prime)\over
16}\left[\Im\psi_i(t^\prime)
{\partial\over\partial\epsilon_1}-\Re\psi_i(t^\prime)
{\partial\over\partial\epsilon_2 }\right]\left[{\partial^2\over
\partial\epsilon_1^2}+{\partial^2\over
\partial\epsilon_2^2}\right]\Omega_{\rm
cl}(\psi_j^\star(t),\psi_j(t),\epsilon_1,\epsilon_2).
\label{quant_corr}
\ee
\end{widetext}

Numerically both the leading term $\langle \Omega(t)\rangle_0$ and
the next correction $\langle \Omega(t)\rangle_1$ are evaluated using
Monte-Carlo integration schemes. The third order derivatives in
Eq.~(\ref{quant_corr}) are found using finite differences, e. g.
\beq
&&{\partial^3 \Omega(\epsilon_1)\over\partial \epsilon_1^3}\approx
{\Omega(2\epsilon_1)-\Omega(-2\epsilon_1)-2\Omega(\epsilon_1)+2\Omega(\epsilon_1)\over
2\epsilon_1^3}\\
&& {\partial^3 \Omega(\epsilon_1,\epsilon_2)\over\partial
\epsilon_1\partial\epsilon_2^2}
 \approx {1\over 2\epsilon_1\epsilon_2^2 }\biggl(\Omega(\epsilon_1,\epsilon_2)+\Omega(\epsilon_1,-\epsilon_2)\\
 &&-\Omega(-\epsilon_1,\epsilon_2)
-\Omega(-\epsilon_1,-\epsilon_2)-2\Omega(\epsilon_1,0)+2\Omega(-\epsilon_1,0)\biggr).\nonumber
\eeq

\begin{figure}[ht]
\includegraphics[width=10cm]{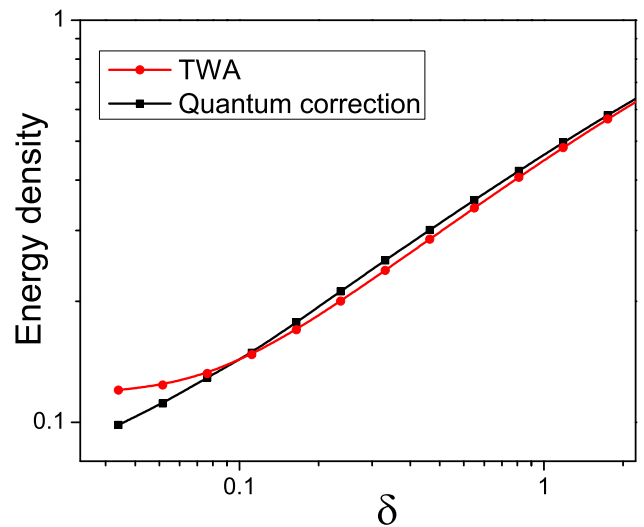}
\caption{ Dependence of the energy density  $\Delta \mathcal E$ on
the $\delta$ at zero temperature with and without the quantum
correction (\ref{quant_corr}). For the details of the calculation
and the parameters of the problem see Fig.~(1) of the main text.
Obviously at small values of $\delta$ the TWA breaks down and one
has to include the correction (\ref{quant_corr}).}
\label{fig:twa_quant}
\end{figure}

It is easy to convince oneself that in order to evaluate these
finite differences one has to simultaneously solve thirteen
Gross-Pitaevskii equations, one for $\epsilon_1=\epsilon_2=0$ and
the others for various combinations of $\epsilon_1, \epsilon_2=0,
\pm\epsilon, \pm 2\epsilon$. While solving thirteen Gross-Pitaevskii
equations is certainly more time consuming task than solving one equation, it
is still tremendously more advantageous than dealing with the exact
quantum problem. To illustrate the importance of quantum correction
at zero temperature we show comparison of dependence $\Delta
\mathcal E(\delta)$ at zero temperature with and without this
correction (see Fig.~\ref{fig:twa_quant}).
The semiclassical approximation gives spurious saturation (and even
increase) of the heating induced in the system as $\delta\to 0$. At
the same time adding the first correction removes this unphysical
behavior and extends the validity of the numerical results to slower
rates.

It is interesting that at finite (even very small) temperatures the
domain of validity of the semiclassical (TWA) approximation
tremendously increases. Indeed the dependence of $\Delta\mathcal E$
on $\delta$ does not show any spurious behavior down to the slowest
rate we were able to analyze (see Figs. 1-3 of the main text). This
result is perhaps intuitively clear: we expect that quantum
corrections play smaller role at higher temperatures. Nevertheless
it is still quite surprising that even very small temperature
$T=0.02$, corresponding to only $1\%$ of the band width $2J$ has
such a strong effect on the validity of the semiclassical making it
virtually exact.

The next very important issue we would like to address here is
whether the Bose-Hubbard model indeed leads to eventual
thermalization and how it affects observables other than energy. We
emphasize that the Bose-Hubbard model is not integrable in all
spatial dimensions and thus thermalization is expected. However, at
low energies the excitations of this model are weakly interacting
long-wavelength phonons. Thus on general grounds one can expect that
the relaxation times of these phonons are very long. We also point
that neglecting relaxation, during our process we primarily populate
low energy excitations, which generically have longer life times
than the high energy excitations. Thus we expect that the relaxation
times in our case will be even longer than in equilibrium.

To analyze thermalization in the system we will concentrate on the
behavior of the correlation functions. According to the
noninteracting theory the long distance behavior of these
correlation functions is given by Eq.~(\ref{eq:30}) if the initial
temperature is zero and by Eq.~(\ref{corr_func1}) at a finite $T$.
\begin{figure}[ht]
\includegraphics[width=9cm]{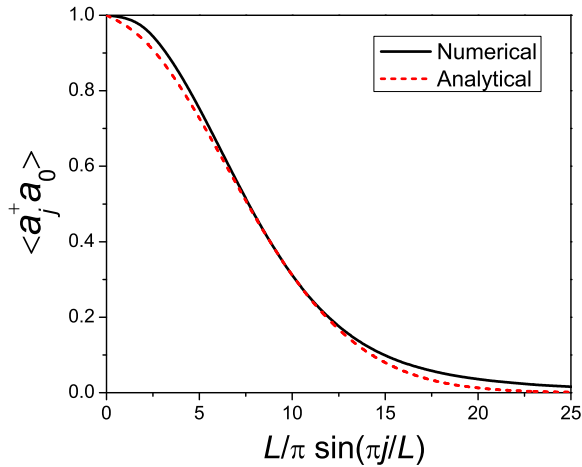}
\caption{Correlation function $\langle a_i^\dagger a_{i+j}\rangle$
of a one-dimensional bosonic Hubbard model as a function of scaled
distance $L/\pi \sin(\pi j/L)$ at $t=3.2/\delta$. The parameters of
the model are identical to those in Fig.~(4) of the main text:
$\delta=0.1$, $L=256$, $T=0.02$. The two lines represent the
numerical data and the analytical result evaluated according to
Eqs.~(\ref{corr_func}), (\ref{sig_eff}), (\ref{alpha_q}), and
(\ref{sig_eff_temp}).}
\label{fig:thermal4}
\end{figure}
In Fig.~\ref{fig:thermal4} we plot correlation functions $\langle
a_j^\dagger a_0\rangle$ at $t=3.2/\delta$ evaluated numerically
(solid black line) and analytically according to
Eqs.~(\ref{corr_func}), (\ref{sig_eff}), (\ref{alpha_q}), and
(\ref{sig_eff_temp}). Because we are dealing with a discrete system
we need to change in all expressions $q\to 2\sin(q_n/2)=2\sin(\pi
n/L)$, where $n$ is an integer. The sum in Eq.~(\ref{corr_func}) is
taken over $n=1,\dots, L-1$. We use $\delta=0.1$, $L=256$, and
$T=0.02$ - the same parameters as in Fig.~(4) of the main text. The
time is chosen such that the interaction $U$ is almost saturated at
$U_0$ and yet the system did not have time to relax to the ground
state. The agreement between the two curves is quite good,
especially given the crudeness of the analytic approximation at this
value of $\delta$, where the heating is significant and the harmonic
approximation to the Hubbard model is not expected to be very
accurate. The small deviation between the two curves can be also due
to partial relaxation of the system by the observation time. We
emphasize that there are no fitting parameters involved in this
comparison. We expect that the agreement between analytic and
numerical results should be even better for smaller values of
$\delta$.
\begin{figure}[ht]
\includegraphics[width=9cm]{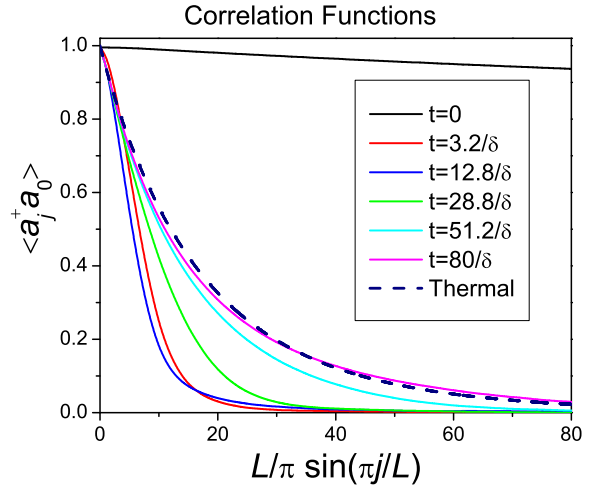}
\caption{Correlation function $\langle a_i^\dagger a_{i+j}\rangle$
of a one-dimensional bosonic Hubbard model as a function of scaled
distance $L/\pi \sin(\pi j/L)$ at different moments of time. Here
$\delta=0.1$, $L=256$, $T=0.02$. This figure duplicates Fig.~4of the main text.}
\label{fig:finte_temp4}
\end{figure}

As we mentioned above the system should eventually thermalize and
the correlation functions should assume the equilibrium form. And
indeed it happens as it is shown in Fig.~4 of the main text~\cite{main}, which
we repeat here for completeness (Fig.~\ref{fig:finte_temp4}). The shape of the correlation function clearly evolves in time and approaches a steady state,
which is very close to the thermal equilibrium. We point out again
that the thermal distribution is obtained for the noninteracting
model with the temperature extracted from the total energy of the
system and thus there are no fitting parameters involved. Obviously
the short distance part of the correlation functions thermalizes
faster than the its long distance tail. This observation is
consistent with general expectations that thermalization times for
short wavelength excitations are shorter. We emphasize that the
overall relaxation time is very long of the order of $10^2-10^3$,
while the natural time scales in the problem, like inverse Josephson
frequency or the inverse frequency associated with steady state
temperature $T\sim 4.8$, are much shorter indicating that the phonon
relaxation times are very long. If one goes to smaller values of
$\delta$ then the thermalization time dramatically increases and the
non-equilibrium shape of the correlation function can be observed
for long times.


\section{Application to cosmology and to adiabatic quantum computation\label{appl}}

\subsection{Small-roll approximation in cosmology. Particle creation.}
The problem of evolution of Universe is inherently adiabatic in
nature. The inflationary cosmology is essentially defined by the
following equations for the scalar field $\phi$ which lives in a
potential $V(\phi)$.
\begin{eqnarray}
\ddot{\phi}+3H\dot{\phi}+\frac{\partial V}{\partial \phi}=0,\qquad
H^2=\frac{8\pi G}{3}\left( \frac{1}{2}\dot{\phi}^2 + V \right),
\end{eqnarray}
It is important that the potential $V=V(\phi)$ has a flat (constant)
part. This allows us to introduce the so-called ``slow-roll''
approximation which can be defined as $\dot\phi^2\ll V(\phi)$,
$|\ddot\phi| \ll |H\dot\phi|$, and $|\ddot\phi| \ll |\partial
V/\partial \phi|$. These approximations imply that the dynamics of
the field $\phi$ is slow, adiabatic. The whole scenario of the
inflationary cosmology is based on this assumption (see e.g.
Ref.~\cite{Linde}). However, as we showed in this paper initial
quantum or thermal fluctuations can be enhanced if the Universe is
in the non-adiabatic regime.

Other potential applications of our findings in cosmology include
the problems of particle creation in the expanding
Universe~\cite{Parker}, the problem which brought a lot of attention
in the literature (see e.g. Ref.~\cite{pc} and references therein).
There, the quantum Hamiltonian of the fluctuating scalar field has
essentially the same form as we used in our manuscript:
\beq
H=\frac{1}{2}\sum_{k}|\partial_{\eta}\phi_{k}|^{2}+\left(k^{2} -\frac{1}{a(\eta)}\partial_{\eta}^{2}a\right)|\phi_{k}|^{2}.
\eeq
Here the conformal time $\eta$ is rescaled by the scaling factor
$a(t)$ which is responsible for the expansion in the
Friedmann-Robertson-Walker metric. The specific effects, which
follow from application of our formalism to these problems remain to
be investigated. We note that recently established connections
 between expansion of the Universe and of the Bose-condensate from a
time-dependent trap~\cite{bec-frw} can be used to experimentally
investigate the effects of non-adiabaticity in the Universe.

As it follows from our work, possible non-adiabatic effects can be
quite significant. In particular, the physics of cosmic microwave
background radiation (CMB) including the prediction of the
temperature is always described using the assumptions of
adiabaticity of the expansion of the Universe. We hope these issues
will be addressed in future by specialists working in cosmology.

\subsection{Adiabatic quantum computation.}
The concept of adiabatic quantum computation was originally proposed
in Ref.~\cite{aqc1} as a method of solving combinatorial
optimization problems. In this approach one starts with a  quantum
Hamiltonian for which the ground state can be easily constructed.
Then the Hamiltonian is adiabatically changed into another one,
whose ground state encodes the solution of the problem. The use of
the adiabatic theorem guarantees that the system will remain in the
instantaneous ground state if the variation of the Hamiltonian is
sufficiently slow. There has been a grown interest in using
adiabatic quantum computation as an architecture for experimental
quantum computation schemes. For practical applications of the
adiabatic quantum computation it is very important whether this
scheme has inherent fault tolerance. Understanding this issue
inspired interest to fundamental questions of the general
applicability of the adiabatic theorem~\cite{aqc2}. Indeed it was
argued that there might be an inconsistency or insufficiency of
conditions in applicability of the adiabatic theorem and for some
specific physical systems. These results further motivated large
amount of works~\cite{aqc3,aqc4,aqc5,aqc6,aqc7,aqc8,aqc9} examining
the applicability of adiabaticity for variety of systems, including
those which are envisioned for quantum computations and cavity QED.
Usually in these works fidelity, i.e. the overlap of the wave
function with the ground state, is used as a measure of
non-adiabaticity. It is possible that real computational schemes can
tolerate small number of excitations in the system. Our analysis
suggests that even this weaker requirement of adiabaticity can be
hard to achieve if regimes {\bf B}) or {\bf C}) are realized.

\section{Evolution of an interacting 1D gas in a time-dependent trap:
Application to the Tonks gas and the Calogero-Sutherland model.\label{TG}}

According to the results of Ref.~\cite{MG} the Tonks gas in an {\it
arbitrary} time dependent parabolic trap
$V_{ext}=m\omega^{2}(t)x^{2}/2$ can be described exactly using the
scaling approach. The evolution of the wave function of TG gas of
$N$ particles is given by the wave function
$\Phi_{TG}(x_{1},\ldots,x_{N};0)$ of the gas at initial time $t=0$
\beq
&&\Phi_{TG}(x_{1},\ldots, x_{N};t)= \frac{1}{b^{N/2}}\Phi_{TG}(x_{1}/b,\ldots,x_{N}/b;0)\nonumber\\
&&~~~~~~~~~~~\times\exp\left(\frac{i\dot{b}}{b\omega_{0}}\sum_{j}\frac{x_{j}^{2}}{2l_{0}^{2}}
-i\sum_{j}E_{j}\tau(t)\right).
\eeq
Here $b(t)$ is the scale factor satisfying the following equation and initial conditions:
\beq\label{eq-b}
\ddot{b}+\omega^{2}(t)b=\omega^{2}_{0}/b^{3},\qquad b(0)=1,\qquad
\dot{b}(0)=0;
\eeq
$l_{0}=\sqrt{\hbar/m\omega_{0}}$ is the oscillator length. We assume
that for $t\leq t_{0}$ the frequency of the trap was fixed at
$\omega=\omega_{0}$ and $E_{j}$ are the single particle energies
corresponding to this frequency. The time parameter $\tau(t)$ is
defined according to $\tau(t)=\int_{0}^{t}dt'/b^{2}(t')$.

Within this approach one can evaluate correlation functions as well
as average energy for time-dependent trap. Here we consider the
time-depending process of "switching off" the trap potential
according to the (relatively general) law
\be\label{omega}
\omega^{2}(t)=\omega_{0}^{2}(1-\delta t)^{r}
\ee
where $r$ is an arbitrary power $0<r<\infty$ and $\delta$ is the
rate of the process. Note that $r=1$ corresponds to the linear ramp
considered in the main paper and $r\to\infty$ with $\delta r$ kept
constant corresponds to the exponential decrease of $\omega^2$ with
time. We will analyze the energy of the system at $t=1/\delta$, i.e.
when the trapping frequency vanishes: $\mathcal E(\delta,r)$. Note
that in the limit $\delta\to 0$ we must have $\mathcal E(0,r)=0$
because the gas can occupy the infinite volume.

\begin{figure}[ht]
\includegraphics[width=9cm]{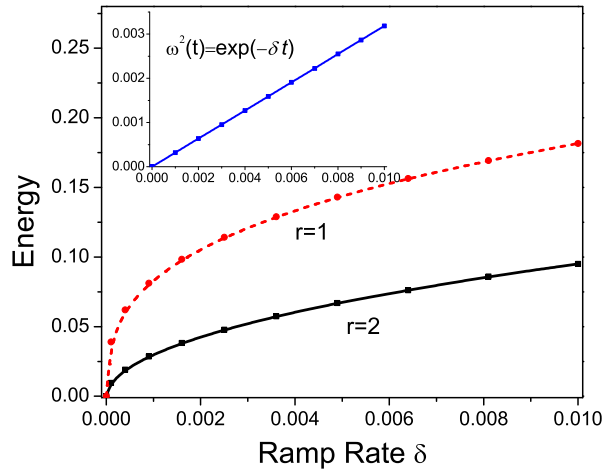}
\caption{Energy of the Tonks gas after release from a trap as a
function of the ramping parameter. The trapping frequency is
changing according to Eq.~(\ref{omega}) with $\omega_0=1$. Shown
curves correspond to $r=1,2,\infty$, where the latter corresponds to
the exponential decrease of $\omega^2$ with time. All three curves
are perfectly fitted by the power law dependence $\mathcal E\propto
\delta^{r/(r+2)}$, i.e. $|\delta|^{1/3}$, $\delta^{1/2}$, and
$\delta$ for the three different curves respectively.}
\label{fig:tonks1}
\end{figure}

Using the scaling approach we described above one can relate the
non-equilibrium energy density to the equilibrium one via
\beq
\mathcal{E}(t,\delta,r)=\frac{1}{2}\mathcal E_0\left[(1-\delta
t)^{r}b^{2}(t)+b^{-2}(t)+[\dot{b}(t)]^{2}\right]
\eeq
where $\mathcal E_0$ is the initial energy of the Tonks gas in the trap and
$b(t)$ is a solution of the Eq.~(\ref{eq-b}) with $\omega^{2}(t)$
given in Eq.~(\ref{omega}). For simplicity we take $\omega^{2}_{0}=1$. The equation for $b(t)$ with
$\tilde{\omega}^{2}(t)=(1-\delta t)^{r}$ can be solved
analytically using the Ermakov approach~\cite{Ermakov}: the solution
of the equation for $b(t)$ is given by
\beq
b(t)=\sqrt{x_{1}^{2}(t)+x_{2}^{2}(t)}
\eeq
where $x_{1,2}(t)$ are solutions of the linear Hill-type equations
\beq
\ddot{x}_{1}+\tilde{\omega}^{2}(t)x_{1}(t)&=&0,\quad
x_{1}(0)=1,\quad\dot{x}_{1}(0)=0,\label{x1}\\
\ddot{x}_{2}+\tilde{\omega}^{2}(t)x_{2}(t)&=&0,\quad
x_{2}(0)=0,\quad\dot{x}_{2}(0)=1\label{x2}
\eeq
These equations can be further solved in terms of the Bessel
functions. However, their explicit form is rather cumbersome and we
rather not show them here. We found that the the energy the rate
$\delta$ satisfies the following scaling
\beq
\mathcal{E}(\delta,r)\approx \omega_{0} C(r)
\left|\frac{\delta}{\omega_{0}}\right|^{\frac{r}{r+2}},
\label{er}
\eeq
where $C(r)$ is a number of the order of unity. Note that at
$r\to\infty$ the exponent in the power of $|\delta|$ saturates at
unity, i.e. the system remains in the regime {\bf B}). In
Fig.~\ref{fig:tonks1} we illustrate the scaling~(\ref{er}) for
several values of $r$.

\begin{figure}[ht]
\includegraphics[width=9cm]{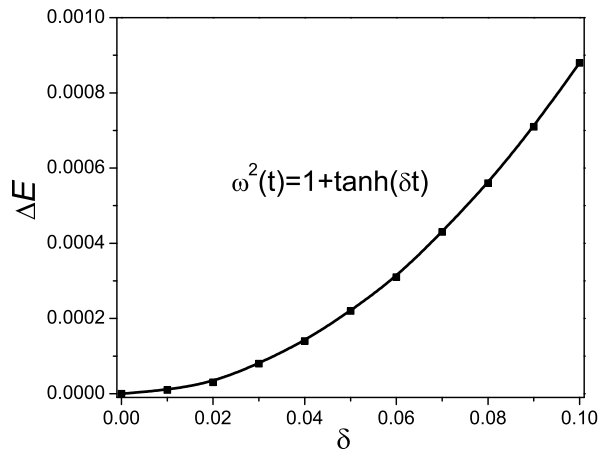}
\caption{Excess energy of the Tonks gas after ramping on the trap
frequency according to $\omega^2(t)=1+\tanh(\delta t)$ as a function
of $\delta$. The dependence perfectly agrees with the quadratic law
$\Delta \mathcal E\propto \delta^2$, i.e. this process belongs to
the analytic ({\bf A}) regime according to our classification.}
\label{fig:tonks2}
\end{figure}

Thus, according to the classification scheme of the main text~\cite{main}, the
TG gas adiabatically released from the harmonic trap follows regime
{\bf B}). We note that in the case when the harmonic trap is not
switched off completely or if one considers the process where the
trapping frequency increases in time the energy dependence on
$\delta$ becomes quadratic. So according to our classification the
system is then in the analytic {\bf A}) regime. In
Fig.~\ref{fig:tonks2} we show the corresponding dependence of
$\Delta\mathcal E(\delta)=\mathcal E(\delta)-\mathcal E(0)$ on
$\delta$ at $t\to\infty$ for the trapping frequency increasing in
time: $\omega^2(t)=1+\tanh(\delta t)$.
Note that not only the dependence is quadratic the heating is almost
negligible compared to what one gets in the {\bf B}) regime (see
Fig.~\ref{fig:tonks1}).

It turns out that the analysis above can be immediately generalized
to another very well known Calogero-Sutherland model describing
one-dimensional Fermions interacting via $1/x^2$
potential~\cite{CS}. This model is a rear example of solvable models
which describe long-range interacting quantum systems. As an
effective model it received a number of applications in various
fields including quantum Hall effect, random matrix theory, etc. In
a time-depending harmonic potential the corresponding time-dependent
Schr\"{o}dinger equation describing assumes the form:
\beq\label{CS}
i\frac{\partial\Psi}{\partial t} &=&\biggl(
-\frac{1}{2}\sum_{j=1}^{N}\frac{\partial^{2}}{\partial x_{j}^{2}}\nonumber\\
&+&\sum_{j>
i=1}^{N}\frac{\lambda(\lambda-1)}{(x_{i}-x_{j})^{2}}+\frac{\omega^2(t)}{2}\sum_{j=1}^{N}x_{j}^{2}\biggr)\Psi
\eeq
Using the scaling ansatz similar to the one employed for the Tonks
gas, the particular solution of this equation, which corresponds to
the ground state at $t=0$, can be written as~\cite{CS2}
\beq
\Psi(\{x_j\},t)={1\over b^{N/2}}e^{i \frac{\dot b}{b}\sum_{j=1}^{N}x_{j}^{2}}\prod_{j\geq
i=1}^{N}\left({|x_{j}-x_{i}|\over b}\right)^{\lambda}\nonumber\\
\eeq
where the function $|b|$ satisfies Eq.~(\ref{eq-b}) with the same
initial conditions. We again assumed $\omega^2(0)=1$ for simplicity.
Note that there is close analogy between the scaling approach for
the Tonks and Calogero-Sutherland models coming from the fact that
the interaction energy in the latter scales in the same way with $x$
as the kinetic energy. Similarly the "equipartition theorem" is
satisfied for the model~(\ref{CS}) in equilibrium as well as for the
TG gas: the half of the total energy comes from the harmonic well.
This immediately implies that the scaling of the energy with
$\delta$ for the adiabatic turning off the potential is identical
for the two models. Thus we conclude that for the dependence
$\omega^2=(1-\delta t)^{r}$ the residual energy at $t=1/\delta$
again scales as $|\delta|^{r/(r+2)}$ and the model belongs to the
class {\bf B}) according to our classification. It is interesting
that this conclusion (which is valid only for diagonal correlations,
e.g. for the energy) does not involve the dependence on the
parameter $\lambda$.

\section{Dicke model\label{Dicke}}

\begin{figure}[ht]
\includegraphics[width=9cm,angle=0]{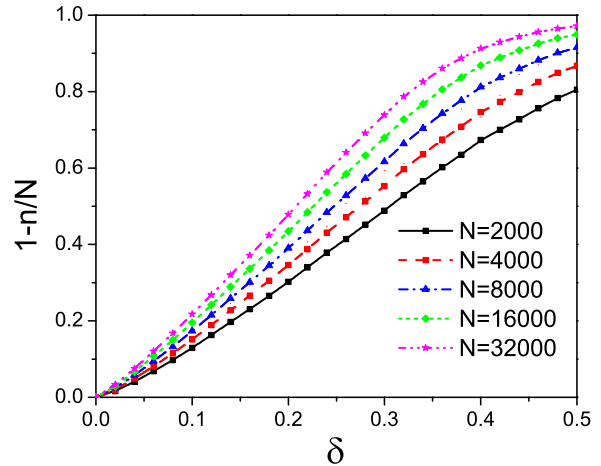}
\caption{Bosonic occupation number for the Dicke model as a function
of driving parameter $\delta$ for different values of total number
of particles $N$. The data is read off after sufficiently large
evolution time. }
\label{dicke}
\end{figure}

In this section we briefly consider a Dicke model, described by the
following Hamiltonian:
\beq
H_{Dicke}=\mu(t) a^{\dag}a+{g\over \sqrt{N}}\sum_{i=1}^{N}\left[a^{\dag}S_{i}^{-}+a S_{i}^{+}\right],
\eeq
where $a$ and $a^\dagger$ are the bosonic fields and $S^{+}$ and
$S^-$ are the bosonic spin-1/2 razing an lowering operators. This
Hamiltonian appears in many contexts of condensed matter, atomic and
optical physics. For example, this model represents interaction of
$N$ two level system with a photon cavity mode. This model also
describes Feschbach resonance of interacting fermions in the case of
a broad resonance~\cite{GA}.

As a particular example we assume that $\mu(t)=-2\delta t$ linearly
changes in time and that initially (at $t\to -\infty$) all spins are
aligned along the $z$ direction and the bosonic mode is empty. This
setup is identical to that considered recently by A.~Altland and
V.~Gurarie~\cite{GA}. In the limit $\delta\to 0$ one expects that
the system will follow the ground state and all spins will flip so
that the population of the bosonic mode $n=\langle a^\dagger
a\rangle$ at $t\to\infty$ is exactly $N$. In Ref.~[\onlinecite{GA}]
it was indicated, however, that one approaches this limit in a
nontrivial way.

In Fig.~(\ref{dicke}) we plot the numerically found dependence
$1-n/N$ as a function of the parameter $\delta$ for various values
of $N$. We were using a semiclassical approach similar to the one
described in Appendix~\ref{App:C}. Let us first  note that the
dependence of $n$ on $\delta$ is linear in agreement with the
regimes {\bf B}) or {\bf C}) in our classification scheme. Second we
observe that there is no adiabatic limit for $N\to\infty$ suggesting
that in fact the dynamics belongs to the regime ({\bf C}). We will
present a more detailed analysis of the slow dynamics of the Dicke
model in a separate publication~\cite{pg}.

{\em Acknowledgements} Correspondence and requests for materials
should be addressed to A.P. We would like to acknowledge  E.~Altman,
E. Demler, A.~Garkun, S.~Girvin, V. Gurarie, M.~Lukin, V.~Pokrovsky,
and N. Prokof'ev for useful discussions. A.P. was supported by AFOSR
YIP and partially by NSF under Grant PHY05-51164. V. G. is partially
supported by the Swiss National Science Foundation and AFOSR. A.P.
also acknowledges Kavli Institute for Theoretical Physics for
hospitality.


\begin{thebibliography}{99}

\bibitem{main} Polkovnikov, A. and Gritsev, V. Breakdown of the adiabatic limit in low dimensional gapless systems, arXiv:0706.0212, to appear in Nature Physics.

\bibitem{KWW}
Kinoshita, T., Wenger, T., and Weiss, D. S. Observation of a One-Dimensional Tonks-Girardeau Gas,  Science {\bf 305}, 1125 (2004).

\bibitem{CS}
Calogero-Moser-Sutherland Models, CRM Series in Mathematical
Physics, Eds. J. F. van Diejen, and L. Vinet (Springer, 2000).

\bibitem{Dicke-model} Dicke, R. H., Coherence in Spontaneous Radiation Processes, Phys. Rev. {\bf 93}, 99 (1954).

\bibitem{adiabatic} Polkovnikov, A. Universal adiabatic dynamics in the
vicinity of a quantum critical point. Phys. Rev. B. {\bf 72},
161201(R) (2005).

\bibitem{ps} Pethick, C.~J. and Smith,~H.{\em Bose Einstein Condensation in
Dilute Gases} (Cambridge University Press, Cambridge, 2003).

\bibitem{krishnendu} Sen, D., Sengupta, K., Mondal, S. Defect production in non-linear quench across a quantum critical point, arXiv:0803.2081.

\bibitem{subir} Sachdev, S. {\em Quantum Phase Transitions}
(Cambridge University Press, Cambridge, 1999).

\bibitem{zurek1} Zurek, W.~H., Dorner, U., and Zoller~P. Dynamics of a
quantum phase transition.  Phys. Rev. Lett. {\bf 95}, 105701 (2005).

\bibitem{jacek} Dziarmaga, J. Dynamics of a quantum phase transition:
Exact solution of the quantum ising model. Phys. Rev. Lett. {\bf
95}, 245701 (2005).

\bibitem{fubini} Fubini, A., Falci, G., and Osterloh, A. Robustness
of adiabatic passage through a quantum phase transition. New Journal
of Physics {\bf 9}, 134 (2007).

\bibitem{walls} Walls,~D.~F. and Milburn,~G.~J. {\em Quantum Optics} (
Springer-Verlag, Berlin, 1994).

\bibitem{gardiner} Gardiner~C.~W. and Zoller~P., {\em Quantum
Noise} (Springer-Verlag, Berlin, 2004).

\bibitem{twa} Polkovnikov, A. Quantum corrections to
the dynamics of interacting bosons: Beyond the truncated Wigner
approximation. Phys. Rev. A {\bf 68}, 053604 (2003).

\bibitem{psg} Polkovnikov,~A., Sachdev,~S.,and Girvin,~S.~M.
Nonequilibrium Gross-Pitaevskii dynamics of boson lattice models.
Phys. Rev. A {\bf 66}, 053607 (2002).

\bibitem{kamenev} Kamenev, A.``Keldysh and Doi-Peliti Techniques for Out-of-
Equilibrium Systems'', in {\em Strongly Correlated Fermions and
Bosons in Low-Dimensional Disordered Systems}, ed. by I. V. Lerner
{\em et. al.} (Kluwer Academic Publishers, Dordrecht, 2002), pp.
313–340.

\bibitem{steel} Steel~M.~J., Olsen,~M.~K., Plimak,~L.~I., Drummond,~P.~D., Tan,~S.~M.,
Collett,~M.~J., Walls~D.~F., and Graham~R. Dynamical quantum noise
in trapped Bose-Einstein condensates. Phys. Rev. A {\bf 58},
4824-4835 (1998).

\bibitem{ap_cat} Polkovnikov, A. Evolution
of the macroscopically entangled states in optical lattices. Phys.
Rev. A {\bf 68}, 033609  (2003).

\bibitem{Linde}
Linde, A., Inflationary Cosmology,  arXiv:0705.0164; Linde, A.,
Particle Physics and Inflationary Cosmology, Contemp. Concepts Phys.
{\bf 5}, 1 (2005); Mukhanov, V., Physical Foundations of Cosmology,
Cambridge University Press (2005).

\bibitem{Parker}
Parker,L., Quantized Fields and Particle Creation in Expanding
Universes. I, Phys. Rev. {\bf 183}, 1057 (1969).

\bibitem{pc}
We note here only a few recent papers, in which formalism similar to
one used by us has been applied for the cosmological problems: A. L.
Matacz, Coherent state representation of quantum fluctuations in
early Universe, Phys. rev. D {\bf 49}, 788 (1994); L. P. Grishchuk,
Y. V. Sidorov, Squeezed quantum states of relic gravitons and
primordial density fluctuations, Phys. Rev. D {\bf 42}, 3413 (1990);
D. Campo, and R. Parentani, Inflationary spectra and violations of
Bell inequalities, Phys. Rev. D {\bf 74}, 025001 (2006); D. campo,
and R. Parentani, Inflationary spectra, decoherence, and two-mode
coherent states, Int. J. Theor. Phys. {\bf 44}, 1705 (2005); A. M.
de M Carvalho, C. Furtado, and I. A. Pedrosa, Scalar fields and
exact invariants in a Friedmann-Robertson-Walker spacetime, Phys.
Rev. D {\bf 70}, 123523 (2004).

\bibitem{bec-frw}
Jain, P., Weinfurtner, S., Visser, M., and Gardiner, C. W., Analogue
model of a FRW universe in Bose-Einstein condensatee: application of
the classical field method, arXiv:0705.2077 (2007); Herring, G.,
{\it et.al.} From Feshbach-Resonance Managed Bose-Einstein
Condensates to Anisotropic Universes: Some Applications of the
Ermakov-Pinney equation with Time-Dependent Nonlinearity,
arXiv:cond-mat/0701756 (2007).

\bibitem{aqc1}
Farhi, E., Goldstone, J., Gutmann, S., and Sipser, M., Quantum
Computation by Adiabatic Evolution, arxiv:quant-ph/0001106 (2000).

\bibitem{aqc2}
Marzlin, K. P., and Sanders, B. C., Phys. Rev. Lett. {\bf 93},
160408 (2004); Tong, D. M., Singh, K., Kwek, L.C., and Oh, C. H.,
Phys. Rev. Lett. {\bf 95}, 110407 (2005).

\bibitem{aqc3}
Sarandy, M. S., Wu, L.-A. and Lidar, D. A. Consistency of the
Adiabatic Theorem, Quantum Inf. Process. {\bf 3}, 331 (2004);

\bibitem{aqc4} Pati, A. K. and Rajagopal, A. K. Inconsistencies of the Adiabatic Theorem and the Berry Phase, quant-ph/0405129;

\bibitem{aqc5}  V\'{e}rtesi, T. and Englman, R. Perturbative Analysis of Possible Failures in the Traditional Adiabatic Conditions, quant-ph/0511141;

\bibitem{aqc6} Duki, S., Mathur, H., and Narayan, O. Is the Adiabatic Approximation Inconsistent?, quant-ph/0510131;

\bibitem{aqc7} Comparat, D. General Conditions for Quantum
Adiabatic Evolution, quant-ph/0607118;

\bibitem{aqc8} Jordan, S. P., Farhi, E., and Shor, P. W. Error Correcting Codes For Adiabatic Quantum Computation, Phys. Rev. A {\bf 74}, 052322 (2006);

\bibitem{aqc9} Larson, J. Stenholm, S. Validity of Adiabaticity in Cavity QED, Phys. Rev. A {\bf 73}, 033805 (2006).


\bibitem{MG} Minguzzi, A. and Gangardt, D. M.  Exact Coherent States of a Harmonically Confined Tonks-Girardeau Gas, Phys. Rev. Lett. {\bf 94}, 240404 (2005).

\bibitem{Ermakov}
Ermakov, V. P., Transformation of differential equations, Univ. Izv.
Kiev. {\bf 20}, 1-19 (1880).

\bibitem{CS2}
Sutherland, B. Exact coherent states of a one-dimensional quantum
fluid in a time-dependent trapping potential. Phys. Rev. Lett. {\bf
80}, 3678 (1998).

\bibitem{GA}
Altland, A., Gurarie, V., Many body generalization of the Landau
Zener problem, arXiv:0709.2526.

\bibitem{pg} Gurarie, V., and Polkovnikov, A., to be published.


\end{thebibliography}
\end{document}